\documentclass{article}

\usepackage{arxiv}

\usepackage[utf8]{inputenc} 
\usepackage[T1]{fontenc}    
\usepackage{url}            
\usepackage{booktabs}       
\usepackage{amsfonts}       
\usepackage{nicefrac}       
\usepackage{microtype}      
\usepackage{graphicx}
\usepackage{natbib}
\usepackage{doi}
\hypersetup{
  colorlinks = true,
  urlcolor   = blue,
  linkcolor  = blue,
  citecolor  = blue
}
\usepackage{amsmath}

\title{MARIA: A Framework for Marginal Risk Assessment without Ground Truth in AI Systems}

\author{ {Jieshan Chen$^*$} \\
	CSIRO's Data61, Australia\\
	\texttt{Jieshan.Chen@data61.csiro.au} \\
	\And
	Suyu Ma$^*$\\
	CSIRO's Data61, Australia\\
	\texttt{Suyu.Ma@data61.csiro.au} \\
	\And
	Qinghua Lu \\
	CSIRO's Data61, Australia\\
	\texttt{Qinghua.Lu@data61.csiro.au} \\
	\And
	Sung Une Lee \\
	CSIRO's Data61, Australia\\
	\texttt{Sunny.Lee@data61.csiro.au} \\
	\And
	Liming Zhu \\
	CSIRO's Data61, Australia\\
	\texttt{Liming.Zhu@data61.csiro.au} \\
}

\hypersetup{
pdftitle={A template for the arxiv style},
pdfsubject={xxx},
pdfauthor={xxx},
pdfkeywords={xxx},
}

\begin{document}
\maketitle

\begin{abstract}
Before deploying an AI system to replace an existing process, it must be compared with the incumbent to ensure improvement without added risk. Traditional evaluation relies on ground truth for both systems, but this is often unavailable due to delayed or unknowable outcomes, high costs, or incomplete data, especially for long-standing systems deemed safe by convention.
The more practical solution is not to compute absolute risk but the difference between systems. We therefore propose a marginal risk assessment framework, that avoids dependence on ground truth or absolute risk. It emphasizes three kinds of relative evaluation methodology, including predictability, capability and interaction dominance. 
By shifting focus from absolute to relative evaluation, our approach equips software teams with actionable guidance: identifying where AI enhances outcomes, where it introduces new risks, and how to adopt such systems responsibly.
\end{abstract}


\section{Introduction}
With rapid advances in artificial intelligence (AI), especially large language models (LLMs), organizations increasingly seek to deploy AI systems as replacements or enhancements for existing processes, from business operations to critical fields like healthcare.
Global corporate investment in AI surged to \$252.3 billion in 2024~\cite{aiindexreport}, underscoring not just enthusiasm but urgency. 

\begingroup
\renewcommand{\thefootnote}{\fnsymbol{footnote}}  
\setcounter{footnote}{1}  
\footnotetext{These authors contributed equally to this work.}
\endgroup

A new AI system, however, must prove not only that it performs better than the existing process/system, but also that it does so safely. 
Traditionally, such evaluation involves comparing each system’s outputs to a known ground truth, i.e., the correct or desired result. 
In practice, obtaining ground truth data for real-world processes is often infeasible. Some outcomes are inherently unknowable~\cite{kiyasseh2024framework} or only revealed after long delays (e.g. long-term policy effects). It also can be costly and time-consuming to measure at scale~\cite{levin2025how}. Many incumbent systems have operated for years without systematically collecting labels for ``right'' versus ``wrong'' decisions. 
Even when datasets with nominal ground truth exist, they may be incomplete or uncertain~\cite{lebovitz2021ai}, blindly treating these labels as absolute can be misleading \cite{lebovitz2021ai}. 
These challenges underscore the need for a different approach when comparing an AI system to an incumbent without reliable ground truth.

Therefore, in this paper, instead of focusing on each system’s absolute performance against an elusive ground truth, we argue that evaluators should focus on the \textit{relative difference} between the new AI system and the existing process/system without assuming access to ground truth. In other words, what is the marginal risk introduced or mitigated by adopting the AI? Our proposed framework, \textbf{MA}rginal \textbf{RI}sk \textbf{A}ssessment without Ground Truth (MARIA), shifts the evaluation from absolute outputs to comparative metrics. 
This ground-truth-independent framework consists of three ways to evaluate the relative performance, including predictability, capability, and interaction dominance, each characterized by common evaluative metrics. Specifically, predictability is examined through metrics like consistency, stability, and uncertainty under comparable inputs. Capability is measured through quantitative and reproducible indicators. Interaction dominance is evaluated using game-based evaluation methods that compare relative system behavior.

The contributions of the paper are threefold:
\begin{itemize}
    \item We identify and distinguish four key challenges with ground truth based AI evaluation, including unknowable, delayed, expensive, and withheld ground truth.
    \item We propose a novel ground-truth- and absolute-risk-independent marginal risk assessment framework, named MARIA, which uses relative measures from three kinds of methods, namely predictability, capability and interaction dominance. 
    By concentrating on the delta between an AI system and the existing system/process, the marginal risk assessment framework provides concrete guidance on identifying where AI improves outcomes, where it introduces new risks, and how to adopt AI systems responsibly.
    \item We demonstrate the usefulness of the framework through a real-world use case on document evaluation, showing how marginal risk assessment provides actionable insights in practice.
\end{itemize}

\section{Challenges in Ground Truth based AI Evaluation}

Evaluation methods in AI often assume a single, authoritative ground truth. However, such ground truths may not always exist or may be impractical to obtain. We identify four recurring situations:  

\begin{itemize}
    \item \textbf{Unknowable.} A ground truth is unknowable when real-world consequences cannot be directly observed or traced. For example, a model may score well on toxicity benchmarks, but we cannot reliably infer what level of actual social harm it causes once deployed~\cite{raman2023centering}.  Similarly, in bail decisions, we cannot know what would have happened if detained defendants had been released~\cite{jon2017human}.
    \item \textbf{Delayed.} A ground truth is delayed when it only becomes observable after a long period of time. For instance, the true quality of an approved vendor may not be evident until years later, when issues such as fraud or compliance violations eventually surface.  
    \item \textbf{Expensive.} A ground truth is expensive when it can, in principle, be established but only at prohibitive cost in expert effort, time, or scale. For example, asking three domain experts to each spend five hours reviewing a paper would yield a highly reliable judgment, but doing so for every submission is infeasible in practice.  
    \item \textbf{Withheld.} A ground truth is withheld when it is relatively easy to obtain but not collected or disclosed, often due to unwillingness, privacy concerns, or institutional omission. For instance, organizations may resist tracking human reviewer performance across fairness or accuracy dimensions, even though such evaluation is possible.  
\end{itemize}

Recognizing this spectrum of limitations is essential for designing robust marginal risk assessments. While absolute risk thresholds may seem ideal in theory, they are rarely workable in practice. We therefore emphasize comparative evaluation, focusing on the marginal risk introduced or mitigated by the new system relative to the baseline, rather than relying on absolute metrics grounded in a definitive truth.


\section{MARIA: A FRAMEWORK FOR MARGINAL RISK ASSESSMENT WITHOUT GROUND TRUTH}

\subsection{Formal Definition of Marginal Risks}
We define marginal risks (MR) as the additional risk introduced or the risk mitigated when a new system replaces an existing one:
\begin{equation}
MR = \Delta R = R_{\text{new}} - R_{\text{baseline}},
\end{equation}
where $R$ is a multi-dimensional risk vector encompassing performance, reliability, safety, security, fairness, privacy, compliance, cost, and resilience. Positive values of $\Delta R$ indicate added risk, negative values indicate reduced risk, and zero indicates no change.

\subsection{Baseline Assumptions}
\label{sec:assumptions}
Existing systems, whether human–human, human–machine, or legacy automated workflows, are typically accepted as carrying ``acceptable risk'' because they align with established norms, historical practices, and accumulated expertise. Importantly, this does not mean they are risk-free. Baseline systems often exhibit variability, inconsistency, and even systematic biases, yet such risks are rarely quantified or formally assessed. In practice, outputs are usually judged acceptable because they have not triggered major failures or because no better alternative exists. Therefore, we make core assumptions as below:
\begin{itemize}
    \item There is no ground truth or oracle of risks or performances. 
    \item Expert-based evaluation on proxy metrics or ground truth is expensive, unavailable or unreliable. 
    \item All comparisons rely only on observable outputs and automatically computable metrics. 
    \item Provenance is controlled: if one system has been trained on another’s data, agreement-based methods may be invalid. 
    \item Only a subset of methods may be used depending on data availability, cost, and assumptions. 
\end{itemize}

\subsection{Risk Profile and Proxy Metrics}

This section outlines what to measure when assessing the added risk of an AI system relative to declared baselines, including human-only, prior system, or no system. Results should be reported as differences against each baseline, clearly indicating whether higher or lower values represent improvement. Wherever possible, measures should align with governance priorities such as decision rights, incentives, and accountability, since operational risk is tied to who makes decisions, how they are motivated, and how their actions are recorded and audited.

With the assumptions, marginal risk can be inferred along three complementary dimensions, \textbf{Predictability}, \textbf{Capability}, and \textbf{Interaction Dominance}. Users may select one, two, or all three depending on data availability, context, and validity of assumptions. Together, these dimensions provide a unified, scalable framework for label-free and assumption-aware marginal risk evaluation.

\subsubsection{Predictability}
Predictability measures how stable and internally coherent a system is under repetition, perturbation, and control changes. A system that behaves more consistently and predictably is considered less risky. Common metrics include:
\begin{itemize}
    \item \textbf{Self-consistency:} Dispersion of outputs across multiple runs with identical prompts and parameters. Estimate via repeated trials varying only the random seed, summarizing with average pairwise similarity or intraclass correlation.
    \item \textbf{Cross-consensus:} Measures agreement among multiple instances or sub-models of the same system, or between independent systems responding to identical inputs. High cross-consensus suggests coherent and reliable system behavior, even when no ground truth is available.
    \item \textbf{Input stability:} Invariance of outputs under semantics-preserving transformations such as paraphrase, order shuffle, or redaction. Test by generating controlled variants per input and measuring output similarity (embedding, edit distance, label agreement). 
    \item \textbf{Control stability:} Smooth, monotonic responses to controllable parameters such as temperature or safety level. Probe along one control axis at a time and assess slope and constraint adherence. 
    \item \textbf{Uncertainty governance:} Regularity of entropy, disagreement, or abstention behavior as inputs grow more ambiguous. Evaluate with confidence-accuracy curves, selective risk at fixed coverage, and abstain rates on hard or noisy cases. 
\end{itemize}

Predictability assumes that perturbations or transformations preserve meaning, systems are independent and not trained on each other’s evaluation data, and randomness is properly sampled through multiple seeds. These metrics are derived from quantitative measures such as variance, entropy, and similarity, which can be computed automatically or verified through narrow LLM-as-judge checks (e.g., contradiction or coherence). This makes the approach fully automated, scalable, and human-free for large-scale consistency testing.

\subsubsection{Capability}
Capability represents the most direct and quantitative evaluation strategy.
It assesses the functional competence of the underlying LLM and the system built upon it, under the assumption that higher capability, if left uncontrolled, also implies higher potential for harm.
Common metrics include:
\begin{itemize}
    \item \textbf{Model-level benchmarks:} Measure the generic performance of the underlying LLM on established reasoning, language, coding, or multimodal benchmarks. Stronger backend models' performance at this level indicates higher potential effectiveness, and by extension, higher downstream capability for the system built on top of it.
    \item \textbf{System effectiveness and operational efficiency:} Evaluate both the functional effectiveness (e.g., agreement with human reviewers, accuracy, or task success rate) and the operational efficiency (e.g., latency, throughput, computational cost). Together, these metrics capture how well and how efficiently the system performs its intended task in real-world contexts.
\end{itemize}

This evaluation strategy assumes that capability can be measured under comparable tasks, data, and resource settings, and that selected benchmarks remain automatable and uncontaminated by prior fine-tuning.
Evaluations should follow standardized, reproducible procedures to ensure fair comparison across systems.
Because capability metrics are inherently quantitative and computable, they enable rapid, low-cost assessment of system performance and potential risk exposure associated with increasing functional competence.

\subsubsection{Interaction Dominance}
Interaction Dominance evaluates how systems behave relative to each other in structured, symmetric games where win conditions are simple and automatically scored. It captures behavioral robustness, adaptability, and emergent risks that may arise through interaction rather than isolated task performance.
Common metrics include:
\begin{itemize}
    \item \textbf{Persuasion duels:} measure belief shifts induced in the opponent, with penalties for unjustified self-shifts.
    \item \textbf{Prediction–surprise games:} reward accurate prediction of the opponent’s next move and the generation of novel counterarguments.
    \item \textbf{Compression–reconstruction games:} assess clarity and faithfulness under concise communication constraints.
\end{itemize}

This dimension assumes that win rules are symmetric, computable, and relevant to the assessed risk, that any LLM-as-judge components are validated on narrow scoring tasks, and that systems are independent and not trained on each other’s transcripts or strategies. All outputs are measurable through automatic similarity, contradiction, or novelty checks. These games can be executed in parallel across thousands of matches, producing statistically robust comparative scores that reveal vulnerabilities or adaptation failures not evident in traditional benchmarks.

\subsection{Evaluation Methodology}

\begin{figure*}[t]
  \centering
  \includegraphics[width=\textwidth]{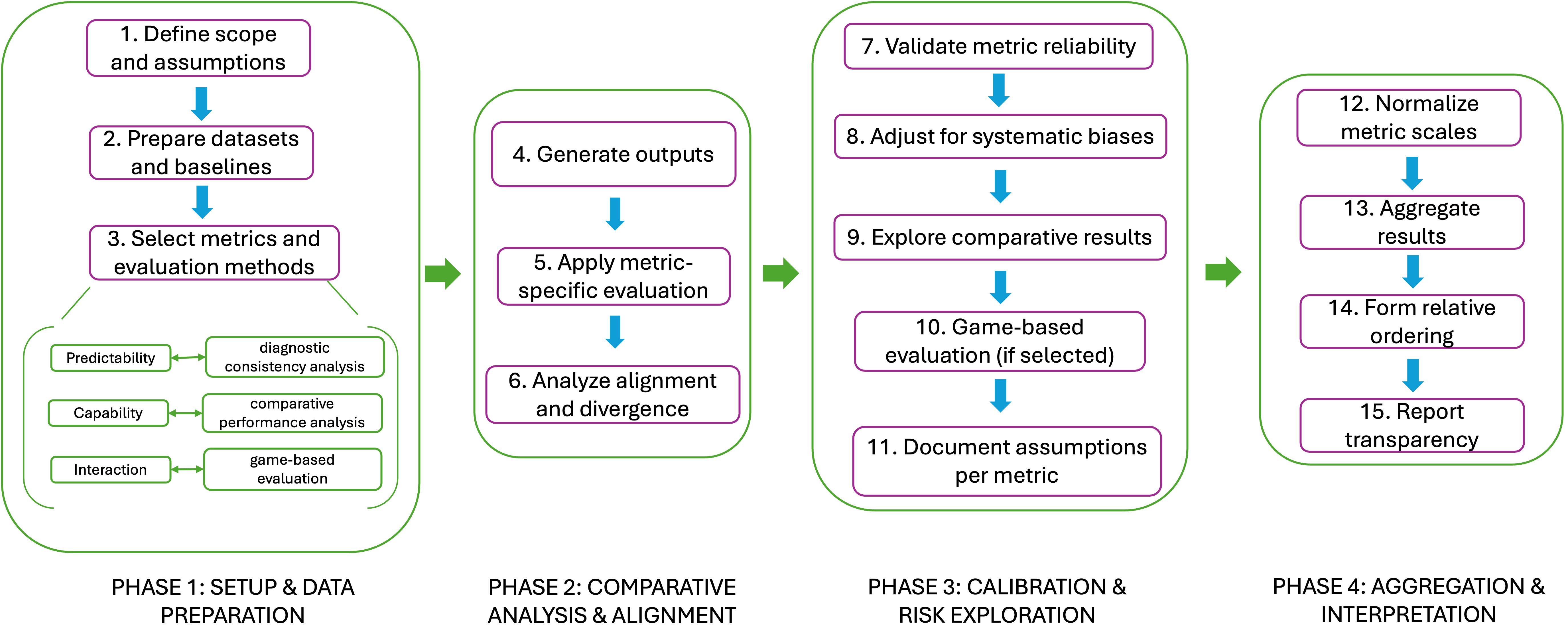} 
  \caption{MARIA workflow.}
  \label{fig:workflow}
\end{figure*}

This section provides a structured, assumption-aware, and scalable process for implementing the MARIA framework.
The workflow operates across four phases, including Setup and Data Preparation, Comparative Analysis and Alignment, Calibration and Risk Exploration, and Aggregation and Interpretation. The overall workflow is illustrated in Fig.~\ref{fig:workflow}.

\subsection*{Phase 1: Setup and Data Preparation}

\textbf{Step 1. Define scope and assumptions.}  
Before conducting analysis, evaluators should clarify the evaluation context and explicitly acknowledge that ground truth or expert labels are unavailable. They should also state explicit assumptions such as monotonicity, system independence, and data provenance to ensure the evaluation remains interpretable and valid.

\textbf{Step 2. Prepare datasets and baselines.}  
Evaluators should assemble datasets that cover representative and realistic use cases, and identify baselines or reference systems, such as human-only or prior AI models, for comparative evaluation. This ensures that baselines are valid and reflect meaningful operational benchmarks.

\textbf{Step 3. Select metrics and evaluation methods.}  
Evaluators should determine which dimensions are appropriate for the task, which specific metrics within each are feasible and cost efficient, and which assumptions are likely to hold. This ensures that evaluation remains efficient, interpretable, and assumption-aware. There is no need to apply all dimensions or all metrics simultaneously. In practice, when human and AI evaluators disagree, scarce ground truth checks should be reserved for critical or high-risk cases only, since even human–human agreement rates are limited. This selective verification approach reduces expert labeling effort while focusing validation where it matters most.

Evaluators should first select one or more evaluation methods suited to the context, and within each chosen method, identify appropriate metrics and map them to compatible evaluation procedures:
\begin{itemize}
  \item \textbf{Predictability metrics} (e.g., self-consistency, input stability, uncertainty governance) → apply \textit{diagnostic consistency analysis}, focusing on variance, entropy, or repeatability.
  \item \textbf{Capability metrics} (e.g., benchmark scores, operational indicators, controllability measures) → apply \textit{comparative performance analysis}, focusing on distributional alignment and systematic bias detection.
  \item \textbf{Interaction metrics} (e.g., persuasion duels, prediction–surprise, compression–reconstruction) → apply \textit{game-based  evaluation}, where outcomes emerge from structured interactions between systems.
\end{itemize}
Each metric–method pair defines a localized evaluation path. Automated judges (embedding similarity, contradiction checkers, or rule-based scores) are assigned as appropriate. Consistent random seeds, budgets, and configurations are set to ensure reproducibility.

\subsection*{Phase 2: Comparative Analysis and Alignment}

\textbf{Step 4. Generate outputs.}  
Evaluators should run each system using identical inputs and configurations, and collect outputs, metadata, and log probabilities, ensuring all runs are reproducible and traceable.

\textbf{Step 5. Apply metric-specific evaluation.}  
Evaluators should execute the evaluation methods chosen in Step 3, tailored to each metric type:
\begin{itemize}
  \item For predictability metrics, assess variance and self-consistency under repeated or perturbed conditions.
  \item For capability metrics, measure distributional shifts, mean differences, and calibration offsets.
  \item For interaction metrics, compute win rates or strategy diversity from game-based experiments.
\end{itemize}

\textbf{Step 6. Analyze alignment and divergence.}  
Evaluators should aggregate results across metrics to identify systematic divergence or hidden correlations. They should examine whether observed differences stem from model instability, capability imbalance, or interactive dominance effects, and focus subsequent steps on the most informative metrics, those showing high disagreement or variance.

\subsection*{Phase 3: Calibration and Risk Exploration}

\textbf{Step 7. Validate metric reliability.}  
Evaluators should assess metric stability by running repeated trials, check metric variance and confirm that automated judges behave consistently on simple control tasks, such as paraphrased or reordered inputs, to ensure reliability.

\textbf{Step 8. Adjust for systematic biases.}  
Biases introduced by sampling strategies or metric design should be identified and corrected. Calibration techniques such as scaling, monotonic transformation, or quantile mapping can be applied to align distributions while preserving interpretability.

\textbf{Step 9. Explore comparative results.}  
Evaluators should analyze areas of both agreement and disagreement to reveal behavioral differences between systems. Special attention should be given to novel or unstable cases where outputs diverge strongly, as these regions often correspond to high marginal risk or untested capability boundaries.

\textbf{Step 10. Game-based evaluation.}  
If Interaction Dominance was selected, evaluators should run targeted evaluation games on disagreement regions or edge cases. Use controlled, symmetric setups (e.g., persuasion duels, prediction–surprise games, or compression–reconstruction tasks) to probe robustness and adaptability. Observed strategies, vulnerabilities, and safety breaches should be documented to inform future risk mitigation.

\textbf{Step 11. Document assumptions per metric.}  
For every metric and comparison, evaluators should record the assumptions required for interpretation and note whether they held empirically. This ensures traceability and transparency in downstream reporting.

\subsection*{Phase 4: Aggregation and Interpretation}

\textbf{Step 12. Normalize metric scales.}  
Evaluators should convert all computed metrics to directional scores so that higher values consistently indicate greater stability or performance. This standardization ensures cross-metric comparability.

\textbf{Step 13. Aggregate results.}  
Evaluators should combine results across dimensions using weighting schemes or rank-based aggregation (e.g., Bradley–Terry or Copeland methods). Uncertainty intervals and sensitivity analyses should be reported to clarify how conclusions depend on metric selection.

\textbf{Step 14. Form relative ordering.}  
Evaluators should determine system dominance by declaring that one system dominates another only if it performs no worse on any selected dimension and strictly better on at least one. If results conflict, mark the systems as incomparable and specify which additional tests or assumptions could resolve the ambiguity.

\textbf{Step 15. Report transparency.}  
Evaluators should provide a full record of assumptions, selected metrics, judges, configurations, and normalization methods. This documentation supports reproducibility, auditability, and interpretability for future evaluations.

\section{Case Study}
Document evaluation is a routine and indispensable part of organizational workflows. Governments must regularly assess tender documents, grants, and proposals to decide which organizations and topics to support. Similarly, the research community carries a continual reviewing workload for funding requests and for conference and journal submissions. Traditionally, such evaluation relies on two or more reviewers who provide judgments based on predefined criteria.

However, with the growth of the knowledge economy, increasing societal complexity, and the rise of new technologies, the demand for high-quality reviews has begun to exceed available human resources. Even a small fraction of review tasks can consume significant time and effort, leaving staff with insufficient capacity for higher-priority work~\cite{aczel2021billion}. Against this backdrop, we use document evaluation as a use case to illustrate the usefulness of the MARIA framework.

The case study is scoped to Phases 1 and 2 of the MARIA framework to demonstrate its core analytical procedures.
Later phases will be implemented in future extensions.

\subsection{Scenario Description}

An organization is responsible for evaluating applications based on the textual materials submitted by applicants. Each application is independently reviewed by two human evaluators, who score it across several predefined dimensions on a 1-to-5 scale. Each dimension is assigned a weight, and a final weighted score is computed for each evaluator. If the difference between the two weighted scores exceeds a threshold, say one full point, a third human evaluator is triggered to resolve the discrepancy. 

Now the team proposes to introduce an AI system, such as an LLM prompted with the same evaluation criteria, to replace one of the human evaluators. The central question arises: \textit{Is the new human–AI system or AI only system riskier than the existing human–based system?}

\subsection{Evaluation Challenges}
Answering this question cannot rely on measuring the absolute correctness or absolute risk of either system. No ground truth exists to determine whether historical human scores or final outcomes were ``correct.'' Even when outcomes are observed (e.g., an application was accepted or rejected), they are shaped by contextual and policy factors rather than universal correctness. Furthermore, stakeholders in the current system may resist any evaluation that challenges the validity of existing human-based systems.

\subsection{Assumptions for Marginal Risk Assessment}
To enable feasible evaluation, we adopt the MARIA framework, which focuses on the relative differences between new and baseline systems. Three assumptions guide this approach:  
\begin{enumerate}
  \item The human–human system is accepted as a baseline with tolerable levels of risk and performance.  
  \item The AI system has not been trained on use case–specific historical data in ways that replicate existing risks.  
  \item The AI does not introduce qualitatively new risks that overwhelm existing safeguards.  
\end{enumerate}

\subsection{Applying the MARIA Framework}

To assess how the risk profile changes when AI reviewers are introduced, we applied both \textit{predictability-based} and \textit{capability-based} evaluation methods.  
Each method was instantiated through quantitative proxies tailored to the application review scenario.

\paragraph{Predictability-based Evaluation.}
This analysis examines whether the system behaves consistently under repetition and controlled perturbations.  
We conducted two tests: (i) \textbf{Self-consistency}, by re-evaluating each application ten times under identical prompts and measuring the variance of normalized scores across runs;  
(ii) \textbf{Input stability}, by comparing outputs for paraphrased and reordered inputs using embedding-based similarity.
All metrics were computed automatically without human annotation.

\paragraph{Capability-based Evaluation.}
This evaluation focuses on system-level functional effectiveness and operational efficiency, which is assessed through:
\begin{itemize}
  \item \textbf{Performance:} Agreement between human–AI and human–human reviewer pairs, indicating consistency in decision quality.
  \item \textbf{Final decision consistency:} Stability of overall outcomes after reconciliation or aggregation, measured by the frequency of third-review triggers.
  \item \textbf{Fairness:} Shifts in decision distributions across document types or applicant groups, revealing potential biases.
  \item \textbf{Security:} Emergence of new vulnerabilities, such as prompt injection, adversarial inputs, or data leakage, which are unique to AI-enabled workflows.
  \item \textbf{Efficiency:} Differences in review time and throughput between configurations, capturing productivity and resource impacts.
\end{itemize}

\subsection{Findings}
\textit{Predictability.}
Repeated evaluations showed that both human and AI reviewers produced stable results across runs, indicating high self-consistency.
However, systematic level differences were observed between the two, suggesting that while each system is internally reliable, their scoring behaviors are not directly interchangeable.
Under paraphrased and reordered inputs, scores remained largely consistent, though the AI system exhibited slightly higher sensitivity to input variation.

\textit{Performance.}
Human–AI agreement was moderate and generally lower than human–human agreement, with discrepancies concentrated in specific criteria.
These variations highlight the need for targeted calibration where alignment remains weak.

\textit{Final Decision Consistency.}
The integration of an AI reviewer altered the frequency of third-review triggers, indicating a shift in reconciliation workload and decision stability.
Monitoring these triggers provides a practical signal for assessing process-level risk.

\textit{Fairness.}
Preliminary analysis revealed mild distributional shifts that may advantage large vendors under capacity-weighted criteria.
Ongoing fairness audits and human oversight are recommended to mitigate potential bias reinforcement.

\textit{Security and Efficiency.}
The AI-assisted workflow introduces new exposure points, such as prompt injection or data leakage, but also improves throughput and reduces manual review time.
These efficiency gains must be balanced against the need for secure input handling, systematic logging, and periodic red-team testing.

Together, these analyses provide a complementary view: predictability methods test the stability and coherence of AI behavior under perturbation, while capability metrics quantify its practical performance, fairness, and resilience when integrated into real-world review workflows.

\section{Discussion}
The marginal risk approach provides a practical way for software teams to understand and manage the extra risk an AI system brings to an existing workflow~\cite{NIST_RMF}. By focusing on risk differences rather than unattainable ground truth, teams can make better decisions about deployment, monitoring, and release.

We recommend including a brief marginal risk summary with every release. This summary should specify the chosen baselines, highlight key risk differences, and report agreement case findings and any calibration steps. Keeping these records concise yet complete will aid transparency and enable ongoing review.
While judge bias, specification gaming, shifting deployment contexts, and evaluation costs remain real challenges, small and repeatable steps, such as adopting a single dialogue game, running agreement case analysis, and publishing a marginal risk summary, can help teams advance AI safety and accountability~\cite{NIST_RMF,Chalamalasetti2023clembench}.

\section{Conclusion}
This paper introduced MARIA, a novel framework for marginal risk assessment without ground truth for evaluating risks introduced by the adoption of AI. Unlike absolute evaluation, which depends on ground truth that may not exist, MARIA highlights the relative risk difference between baseline and new workflows. By formalizing this framing and proposing metrics across multiple dimensions, we show how organizations can assess whether AI adoption increases, decreases, or maintains overall risk. 
Our case study on document evaluation demonstrates how MARIA can be applied in practice, even in domains where correctness is plural or contested. The analysis emphasizes that AI may both reduce certain risks (e.g., reviewer inconsistency) and introduce new ones (e.g., prompt injection or data leakage). Importantly, evaluation should therefore move beyond absolute performance and focus on how the risk profile shifts with AI involvement.

\bibliographystyle{unsrt}
\bibliography{ref}

\begin{thebibliography}{1}

\bibitem{aiindexreport}
{Stanford HAI}.
\newblock Economy |the 2025 ai index report, 2025.
\newblock [ONLINE]. Available: {https://hai.stanford.edu/ai-index/2025-ai-index-report} (URL).

\bibitem{kiyasseh2024framework}
Dani Kiyasseh, Aaron Cohen, Chengsheng Jiang, and Nicholas Altieri.
\newblock A framework for evaluating clinical artificial intelligence systems without ground-truth annotations.
\newblock {\em Nature Communications}, 15(1):1808, 2024.

\bibitem{levin2025how}
{Aaron Levin}.
\newblock How to evaluate your ai product if you don’t have ground truth data, 2025.
\newblock [ONLINE]. Available: {https://www.vellum.ai/blog/how-to-evaluate-your-ai-product-if-you-dont-have-ground-truth-data} (URL).

\bibitem{lebovitz2021ai}
Sarah Lebovitz, Natalia Levina, and Hila Lifshitz-Assaf.
\newblock Is ai ground truth really true? the dangers of training and evaluating ai tools based on experts’ know-what.
\newblock {\em MIS quarterly}, 45(3):1501--1526, 2021.

\bibitem{raman2023centering}
Vyoma Raman, Eve Fleisig, and Dan Klein.
\newblock Centering the margins: Outlier-based identification of harmed populations in toxicity detection.
\newblock {\em EMNLP2023}, 2023.

\bibitem{jon2017human}
Jon Kleinberg, Himabindu Lakkaraju, Jure Leskovec, Jens Ludwig, and Sendhil Mullainathan.
\newblock Human decisions and machine predictions*.
\newblock {\em The Quarterly Journal of Economics}, 133(1):237--293, 08 2017.

\bibitem{aczel2021billion}
Balazs Aczel, Barnabas Szaszi, and Alex~O Holcombe.
\newblock A billion-dollar donation: estimating the cost of researchers’ time spent on peer review.
\newblock {\em Research integrity and peer review}, 6(1):1--8, 2021.

\bibitem{NIST_RMF}
U.S. National~Institute of~Standards and Technology.
\newblock Artificial intelligence risk management framework (ai rmf 1.0).
\newblock {\em NIST Special Publication}, 2023.

\bibitem{Chalamalasetti2023clembench}
Kranti Chalamalasetti, Jana G{\"o}tze, Sherzod Hakimov, Brielen Madureira, Philipp Sadler, and David Schlangen.
\newblock clembench: Using game play to evaluate chat-optimized language models as conversational agents.
\newblock In {\em Proceedings of the 2023 Conference on Empirical Methods in Natural Language Processing}, pages 11174--11219. Association for Computational Linguistics, 2023.

\end{thebibliography}

\end{document}